%% file: main.tex
\documentclass{article}

\usepackage{microtype}
\usepackage{graphicx}
\usepackage{subcaption}
\usepackage{siunitx}
\usepackage{booktabs} 

\usepackage[preprint,nohyperref]{icml2026}
\usepackage{xurl}
\usepackage{hyperref}

\usepackage{amsmath}
\usepackage{amssymb}
\usepackage{mathtools}
\usepackage{amsthm}
\usepackage{listings}
\usepackage{minted}
\usepackage{booktabs}
\usepackage{longtable}
\usepackage{array}
\usepackage{caption}
\usepackage{threeparttable}
\usepackage{booktabs,threeparttable,tabularx,makecell}

\usepackage[capitalize,noabbrev]{cleveref}

\theoremstyle{plain}
\newtheorem{theorem}{Theorem}[section]
\newtheorem{proposition}[theorem]{Proposition}

\theoremstyle{definition}

\theoremstyle{remark}

\usepackage[textsize=tiny]{todonotes}

\icmltitlerunning{Efficient Detection of Bad Items}

\begin{document}

\twocolumn[
  \icmltitle{Efficient Detection of Bad Benchmark Items with Novel Scalability Coefficients}



  \icmlsetsymbol{equal}{*}

  \begin{icmlauthorlist}
    \icmlauthor{Michael Hardy}{equal,yyy}
    \icmlauthor{Joshua Gilbert}{zzz}
    \icmlauthor{Benjamin Domingue}{yyy}
  \end{icmlauthorlist}

  \icmlaffiliation{yyy}{Stanford University, CA, United States}
  \icmlaffiliation{zzz}{Harvard University, MA, United States}

  \icmlcorrespondingauthor{Michael Hardy}{hardym[$\alpha\tau$]stanford[$\cdot$]edu}

  \icmlkeywords{Machine Learning, ICML}

  \vskip 0.3in
]



\printAffiliationsAndNotice{}  

\begin{abstract}
    The validity of assessments, from large-scale AI benchmarks to human classrooms, depends on the quality of individual items, yet modern evaluation instruments often contain thousands of items with minimal psychometric vetting. We introduce a new family of nonparametric scalability coefficients based on interitem isotonic regression for efficiently detecting globally bad items (e.g., miskeyed, ambiguously worded, or construct-misaligned). The central contribution is the signed isotonic $R^2$, which measures the maximal proportion of variance in one item explainable by a monotone function of another while preserving the direction of association via Kendall's $\tau$. Aggregating these pairwise coefficients yields item-level scores that sharply separate problematic items from acceptable ones without assuming linearity or committing to a parametric item response model. We prove that the signed isotonic $R^2$ is extremal among monotone predictors (it extracts the strongest possible monotone signal between any two items) and show that this optimality property translates directly into practical screening power. Across three AI benchmark datasets (HS Math, GSM8K, MMLU) and two human assessment datasets, the signed isotonic $R^2$ consistently achieves top-tier AUC for ranking bad items above good ones, outperforming or matching a comprehensive battery of classical test theory, item response theory, and dimensionality-based diagnostics. Crucially, the method remains robust under the small-n/large-p conditions typical of AI evaluation, requires only bivariate monotone fits computable in seconds, and handles mixed item types (binary, ordinal, continuous) without modification. It is a lightweight, model-agnostic filter that can materially reduce the reviewer effort needed to find flawed items in modern large-scale evaluation regimes.
\end{abstract}

\section{Introduction}
The validity of any assessment--from classroom exams to large-scale AI benchmarks--depends on the quality of its individual items. Even a small number of flawed items (e.g., incorrect answer keys, scoring bugs, ambiguity, construct drift) can distort scores, undermine rankings, and lead to invalid conclusions \citep{casabianca_psychometrics_2025,zhang_how_2025}. In high-stakes human testing, item development is typically accompanied by extensive qualitative review and quantitative pilot analyses, and problematic items are usually removed before operational deployment. By contrast, many modern AI benchmarks are assembled at scale (often synthetically), frequently rely on automated grading for unstructured responses, and may contain thousands of items with minimal psychometric vetting. As a result, benchmark scores can be sensitive to a small set of bad items, and the burden of validation is shifted to downstream users \citep{casabianca_psychometrics_2025,zhang_how_2025,truong_fantastic_2025,reuel_betterbench_2024,salaudeen_measurement_2025}.

This paper focuses on \emph{global bad-item detection}: efficiently prioritizing items for human review when some items are flawed in ways that degrade measurement overall (as distinct from differential item functioning, which concerns group-specific item behavior). The core problem is practical: given a response matrix with many items, how can we rank items so that the worst ones are found quickly?

We propose a new family of nonparametric scalability coefficients based on \emph{interitem isotonic regression}. The main contribution is a signed coefficient derived from the isotonic-regression coefficient of determination, which we call the \emph{signed isotonic $R^2$}. For a response matrix $Y \in \mathbb{R}^{n \times p}$ and for each ordered item pair $(i,j)$, we fit the best monotone relationship between item $i$ and item $j$ and measure the fraction of variance in $Y_j$ explained by a monotone function of $Y_i$. Aggregating these pairwise associations yields item-level scores that sharply separate globally bad items from acceptable ones. Empirically, these coefficients improve practical detection efficiency (measured via AUC for ranking bad items above good ones) and remain computationally scalable for large benchmark regimes.

\section{Background and Motivation}\label{sec:background}
\subsection{What counts as a ``bad item''?}
An item is ``bad'' when it systematically breaks the intended measurement logic of the instrument. Across human assessments and AI benchmarks, we consider four common global failure modes:
\begin{enumerate}
    \item \textbf{Bad key}: the labeled correct answer is wrong.
    \item \textbf{Bad grading}: the scoring procedure is incorrect or inconsistently applied.
    \item \textbf{Ambiguity}: multiple defensible answers or unclear prompt/specification.
    \item \textbf{Construct misalignment}: the item elicits skills outside the intended construct (e.g., format quirks, spurious cues, irrelevant knowledge).
\end{enumerate}
These failure modes typically reduce an item's coherence with the rest of the test, even when the overall instrument is intended to measure a dominant latent trait. Our aim is to detect such \emph{globally} problematic items; we do not address fairness questions or group-conditional anomalies (e.g., DIF\footnote{Items that demonstrate DIF may also be flagged by globally problematic items. Instead of a group differential item functioning, for this study, the reference group is the entire population.}) in this work.

\subsection{Existing tools are often inefficient in large regimes}
A wide range of item-quality indices exist across CTT, IRT, nonparametric scaling \citep{zijlmans_item-score_2018,zijlmans_methods_2018}. These indices differ along three practical axes:

\paragraph{Dynamic: ``bad relative to what?''}
Common comparisons include
\begin{itemize}
    \item \textbf{Interitem} (item vs. item): correlations, agreement, mutual information, pairwise scalability.
    \item \textbf{Item--rest} (item vs. total/others): corrected item--total correlation, item-rest regression.
    \item \textbf{Item-drop} (change in test statistic when removed): $\Delta\alpha$, $\Delta$ reliability.
    \item \textbf{Parameter-based} (item as a member of a model): IRT discrimination/fit, factor loadings.
\end{itemize}

\paragraph{Baseline: ``bad with respect to what property?''}
Different statistics implicitly target different notions of misfit: linear association, mean differences, variance explained, local independence violations, or parameter inconsistency.

\paragraph{Assumptions: ``bad under what model?''}
Many indices are efficient only when their assumptions hold (e.g., linearity, parametric ICC forms, well-behaved latent distributions, sufficient sample size). In AI benchmarks, these assumptions are often strained: item types are mixed (binary/ordinal/continuous scores), grading noise may be structured, and data may be sparse or highly imbalanced. Moreover, some widely used indices are \emph{direction-blind} (they emphasize magnitude but not sign) or collapse nonlinear monotone effects into weaker linear proxies.

\subsection{Item ``behavior'' as interitem social compatibility}
A useful intuition is to treat items as a team: good items ``get along'' with other items that measure the same construct. If a test is approximately unidimensional  responses to any two items should be positively and monotonically related (up to noise and local dependence; \cite{sijtsma_use_2009,revelle_unidim_2025,mair_unidimensional_2015,ten_berge_greatest_2004}). Bad items often exhibit one or more of:
(i) weak association with most other items,
(ii) non-monotone behavior (e.g., middle-ability respondents outperform high-ability respondents due to ambiguity or grading),
(iii) inversions (negative association) consistent with miskeying or systematic scoring reversal.

This motivates \emph{interitem} measures of fit: quantify how well each item participates in the network of expected monotone dependencies.

\subsection{Desiderata for scalable bad-item detection}
We seek item-level indices that are:
\begin{enumerate}
    \item \textbf{Practically efficient}: prioritize bad items early in a review queue.
    \item \textbf{Computationally efficient}: feasible for large $n$ (responses) and very large $p$ (items).
    \item \textbf{Model-agnostic}: avoid reliance on strict parametric IRT forms.
    \item \textbf{Monotonicity-aware}: exploit the key qualitative constraint of unidimensional measurement.
    \item \textbf{Type-flexible}: handle mixed outcome types and asymmetric relationships.
\end{enumerate}

\subsection{Why isotonic regression?}
Isotonic regression provides a principled way to extract \emph{maximal monotone signal} between variables without committing to a parametric functional form. For item analysis, this is attractive because the key expectation under a dominant latent trait is monotonic dependence, not necessarily linear dependence. By measuring how much of an item's variability can be explained by a monotone function of another item, we obtain a natural nonparametric analogue of ``scalability'' that (i) directly targets the monotonicity assumption and (ii) can preserve directionality (through a signed association), enabling sharper detection of inversions such as miskeys.

If a set of items measures a single latent trait, then performance on any two items should be positively and monotonically related. The strength of this monotonic relationship, aggregated across all item pairs, becomes a powerful indicator of item fit. The primary contribution is a new family of nonparametric, nonlinear scalability coefficients built by maximizing the information gained by assuming monotonicity. Taking inspiration from Loevinger’s $H$ \citep{loevinger_systematic_1947},\footnote{The initial inspiration for these proposed solutions.} Mokken scaling \citep{wind_instructional_2017,mokken_theory_2011,sijtsma_monotone_2002,van_der_ark_mokken_2007}, and advances in  isotonic regression in IRT \citep{lee_applications_2002,lee_comparison_2007,lee_use_2009,luzardo_nonparametric_2015,yu_learning_2022}, this approach reframes item fit as a function of its consistent, monotonic behavior with all other items on the scale.

\section{Methods}\label{sec:methods}
\subsection{Notation and setup}
Let $Y \in \mathbb{R}^{n \times p}$ be the response matrix with respondents $r \in \{1,\dots,n\}$ and items $i \in \{1,\dots,p\}$. The column vector for item $i$ is $y_i \in \mathbb{R}^n$. Responses may be binary, ordinal, or continuous; we assume higher values reflect greater success on the construct (after any required recoding). Our goal is to compute, for each item $i$, a scalar \emph{badness score} (or conversely a \emph{fit/scalability score}) used to rank items for review.

\subsection{Interitem isotonic regression}
Fix an ordered pair of distinct items $(i,j)$. We model $y_j$ as a monotone function of $y_i$:
\begin{equation}
    y_{rj} \approx f_{i \rightarrow j}(y_{ri}), 
    \qquad f_{i \rightarrow j} \in \mathcal{F}_{\uparrow},
\end{equation}
where $\mathcal{F}_{\uparrow}$ is the set of non-decreasing functions on the observed support of $y_i$. The isotonic regression estimator is
\begin{equation}
    \hat f_{i \rightarrow j}
    \;\in\;
    \arg\min_{f \in \mathcal{F}_{\uparrow}}
    \sum_{r=1}^n \left(y_{rj} - f(y_{ri})\right)^2,
    \label{eq:isotonic}
\end{equation}
computed efficiently using the Pool Adjacent Violators Algorithm (PAVA; \citep{busing_monotone_2022}) after sorting observations by $y_i$ (with standard handling of ties). This yields fitted values $\hat y^{(i\rightarrow j)}_{rj} = \hat f_{i\rightarrow j}(y_{ri})$.

\paragraph{Asymmetry and mixed item types.}
Because the regression is directional, the strength of $i \rightarrow j$ need not equal $j \rightarrow i$, which is useful when item types differ (e.g., a ordinal partial-credit item may monotonically explain a binary item differently than vice versa). This asymmetry is a feature: it permits detection based on predictable directional structure rather than forcing symmetry.

\subsection{Signed isotonic \texorpdfstring{$R^2$}{R2} as a pairwise scalability coefficient}
We quantify the monotone explanatory power of item $i$ for item $j$ using an isotonic analogue of the coefficient of determination:
\begin{align}
    R^2_{i \rightarrow j}
    &=
    1 - 
    \frac{\sum_{r=1}^n \left(y_{rj} - \hat y^{(i\rightarrow j)}_{rj}\right)^2}
         {\sum_{r=1}^n \left(y_{rj} - \bar y_{j}\right)^2},
    \quad
    \bar y_j = \frac{1}{n}\sum_{r=1}^n y_{rj}.
    \label{eq:r2}
\end{align}
To preserve directionality (inversions) we attach a sign based on the global direction of association between $y_i$ and $y_j$ based on Kendall's $\tau$. Let
\begin{equation}
    s_{ij} = \operatorname{sign}\!\left(\tau(y_i, y_j)\right),
    \label{eq:sign}
\end{equation}
with the convention that $s_{ij}=0$ if the correlation is numerically $0$ or undefined (e.g., zero variance). The \textbf{signed isotonic coefficient} is
\begin{equation}
    \mathcal{M}_{i \rightarrow j}
    =
    s_{ij}\, R^2_{i \rightarrow j}.
    \label{eq:signed_r2}
\end{equation}
Intuitively, $\mathcal{M}_{i\rightarrow j}$ estimates the \emph{proportion of monotone variance explained}, while preserving whether the relationship aligns with the expected positive direction. Bad keys and systematic grading reversals tend to induce negative or unusually small signed values across many pairs.

\subsection{A formal interpretation: optimality among monotone predictors}
The empirical advantage of isotonic $R^2$ is explained by an optimization property: it measures the maximal proportion of variance explainable by \emph{any} monotone transformation.

\begin{proposition}[Maximal monotone explained variance]
\label{prop:max_monotone_r2}
Fix item pair $(i,j)$ and consider predictors of $Y_j$ of the form $f(Y_i)$ where $f$ is non-decreasing. Let $\hat f$ be the isotonic regression solution as defined in Eq. \ref{eq:isotonic}.
Then for any non-decreasing $g$,
\[
\sum_{r=1}^n (y_{rj} - \hat f(y_{ri}))^2 \le \sum_{r=1}^n (y_{rj} - g(y_{ri}))^2,
\]
and therefore $R^2_{i\rightarrow j}$ computed from $\hat f$ is the largest achievable $R^2$ among monotone predictors.
\end{proposition}

\subsection{From pairwise coefficients to item-level badness scores}
For each focal item $i$, we aggregate its pairwise signed isotonic relationships with all other items:
\begin{equation}
    \mathrm{Fit}(i)
    =
    \frac{1}{p-1}\sum_{j \neq i} \mathcal{M}_{i \rightarrow j}.
    \label{eq:item_fit_out}
\end{equation}
We then rank items by increasing $\mathrm{Fit}(i)$ (lower implies more suspicious). Variants we consider in ablations (not required for using the method) include:
(i) symmetrized aggregation $\mathcal{M}_{\text{iso}}=\tfrac{1}{2}(\mathcal{M}_{i\rightarrow j}+\mathcal{M}_{j\rightarrow i})$,
(ii) robust aggregation using trimmed means/medians to reduce sensitivity to local dependence clusters,
(iii) nonnegative aggregation using $|\mathcal{M}_{i\rightarrow j}|$ when direction is known to be unreliable.

\subsection{Computational considerations}
For each ordered pair $(i,j)$, isotonic regression reduces to sorting by $y_i$ and a linear-time PAVA pass. In practice, when many items are binary or low-cardinality, sorting can be implemented via counting/bucketing, making pairwise fits fast. The full pairwise matrix is $O(p^2)$ fits; we therefore use two scalable strategies depending on regime:
\begin{itemize}
    \item \textbf{All-pairs} for moderate $p$ (typical in human assessments).
    \item \textbf{Subsampled neighbors} for very large $p$ (typical in AI benchmarks): compute $\mathrm{Fit}(i)$ using a fixed-size set of comparison items per $i$ (random, stratified by difficulty, or chosen via a computationally cheap pre-screen such as correlation). This preserves ranking quality while reducing compute to $O(pK)$ fits for $K \ll p$.
\end{itemize}

\subsection{Evaluation protocol: efficiency as ranking performance}
We evaluate item-detection \emph{efficiency} by treating each metric as a scoring function that ranks items from most to least suspicious, then computing the area under the ROC curve (AUC) for classifying known bad items. Formally, for item scores $S(i)$ where larger indicates ``more bad'' (we use $S(i)=-\mathrm{Fit}(i)$), AUC equals
\begin{equation}
    \Pr\!\left(S(i_{\text{bad}}) > S(i_{\text{good}})\right),
\end{equation}
the probability that a randomly chosen bad item is ranked above a randomly chosen good item. This directly reflects expected reviewer time saved: higher AUC concentrates bad items earlier in the queue. Thus we are evaluating bad-item detection as a \emph{ranking} problem: a statistic assigns each item $i$ a score $S(i)$, and we sort items from most suspicious to least suspicious. Ground-truth labels $L(i)\in\{0,1\}$ indicate whether an item is globally bad (Sec.~\ref{sec:background}; not DIF). Performance is measured by how effectively the ranking prioritizes bad items for review.

\subsection{Baselines}
To contextualize gains, we compare signed isotonic $R^2$ against a broad suite of established CTT/IRT and association-based indices, covering:
(i) interitem association (e.g., linear and information-theoretic dependence),
(ii) item-rest statistics (e.g., corrected item-total correlation and monotone dependence),
(iii) parameter-based proxies (e.g., discrimination or explained-variance loadings),
(iv) item-drop deltas (e.g., reliability changes).
All methods produce an item ranking, evaluated under the same AUC protocol.

\subsection{Scope}
Our methods target global item misfit detectable via disrupted monotone coherence with the rest of a scale. We do not attempt to diagnose the causal source of misfit (keying vs.\ ambiguity vs.\ construct drift), nor do we conduct group-conditional DIF analyses; rather, we provide a computationally efficient front-end filter that materially improves the rate at which human reviewers find problematic items in both traditional assessments and large AI benchmarks.

\section{Empirical Approach}

\begin{figure*}[htbp]
    \centering
    \includegraphics[width=1\linewidth]{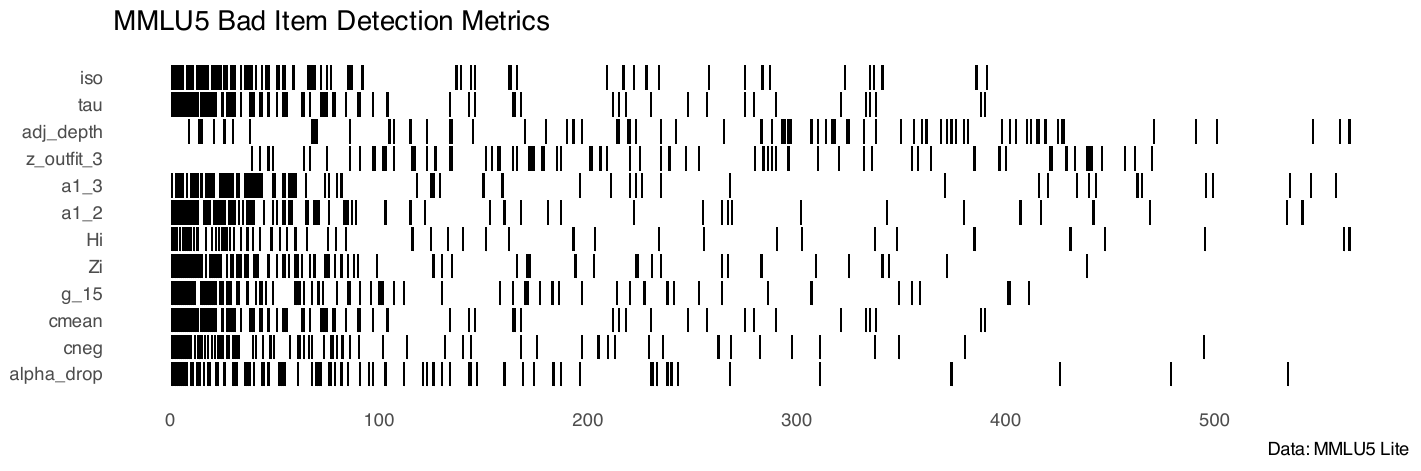}
    \caption{\textbf{Examples of Differences in Item Detection Efficiencies based on Technique}:Bad Item detection as ordered by each item-fit metric. iso = $M_{iso}$; tau = $M_\tau$; adj\_depth = Isolation Forest Tree Depth (anomaly detection), z\_outfit\_3 = standardized absolute 3PL outfit statistic, a1\_3 and a1\_2 = discrimination parameter for 3PL and 2PL, respectively; Hi = $H_i$; Zi = $Z_i$, g\_15 = loading on the general factor of a 15-factor estimation of McDonald's $\omega$; cmean = Mean inter-item tetrachoric correlation; cneg = proportion of inter-item tetrachoric correlations < 0; alpha\_drop = benchmark reliability (Cronbach's $\alpha$) with item removed}
    \label{fig:rugplot}
\end{figure*}

\subsection{Datasets and labels}
\label{sec:datasets_exp}
We use both human assessment data and AI benchmark data.
Human datasets reflect conventional test development pipelines where few bad items survive, but when they do, they are consequential.
AI benchmark datasets represent large item banks assembled without comparable validation pipelines; these are the main motivation for scalable detection tools.

Each dataset includes an externally curated list of globally bad items (bad key, bad grading, ambiguity, construct misalignment). For experiments involving bootstraps, we enforce a minimum number of bad items in each resample so that AUC is well-defined and not dominated by degenerate cases.

\subsubsection{Application Domain: AI Benchmark Assessment}

The emergence of large language models (LLMs) has created unprecedented challenges for psychometric evaluation. AI benchmarks often lack human comparison data, having been designed specifically for machine evaluation. As model performance approaches saturation on existing benchmarks, the identification of problematic items becomes crucial—differences of one or two poor items can determine rankings among state-of-the-art models.

\subsubsection{Unidimensionality Assumption in AI Assessment}

While human cognitive assessment typically reveals multidimensional ability structures, AI models present a unique case. Contemporary LLMs share a fundamental training objective: autoregression of Internet text. Despite subsequent modifications through instruction tuning and reinforcement learning from human feedback (RLHF), we hypothesize that this primary objective dominates performance across diverse benchmarks \citep{mccoy_embers_2023}.

This suggests that for AI evaluation contexts, a unidimensional latent ability model may be appropriate, or alternatively, that the autoregressive ability component substantially outweighs other potential factors. We validate this assumption using the MMLU (Massive Multitask Language Understanding) dataset, which spans five distinct subject areas.

\subsubsection{AI and Benchmarks and Ground Truth}

We employed multiple datasets with independently validated item quality assessments, allowing for objective evaluation of metric performance. Each dataset contained items previously identified as problematic through expert review or statistical flagging procedures. The responses come from Holistic Evaluations of Language Models (HELM) datasets \cite{liang_holistic_2023}.\footnote{\url{https://crfm.stanford.edu/helm/}} From HELM, we utilize the model responses and combined bad item labels of the HS Math, GSM8K, and HELM-Lite 5-subject MMLU Benchmark from the \cite{truong_fantastic_2025, vendrow_large_2025} studies.

\subsubsection{Human Datasets}\label{sec:humdata}
Publicly available datasets datasets with ``bad'' items are rare, as bad items are typically removed before public release. We were fortunate enough to obtain access to two human data, in addition to the main AI benchmark datasets above. , we have two human datasets where ``bad'' items have been identified. One is publicly available and found within the Item Response Warehouse. It has 31 items and 7780 respondents. The second human dataset is a private dataset from an educational intervention pilot study in high school science. The flagged items were identified as either ambiguous or misaligned with the intended construct out of 20 items, each having 106 individual responses. We conjecture that pilot studies, which are rarely accessible publicly, would be a natural format for using these analyses.


\subsection{Evaluation metric: AUC as reviewer-efficiency}
\label{sec:auc}
We operationalize efficiency as the area under the ROC curve (AUC) when metrics are used to rank-order items by suspected quality. AUC represents the probability that a classifier will rank a randomly chosen instance of a bad item the higher than a randomly chosen instance from items without issues. For any item scoring rule $S$, we compute the area under the ROC curve (AUC) for classifying items using $S(i)$:
\begin{align}
\mathrm{AUC}(S)=
\Pr\!\left(S(i_{\text{bad}}) > S(i_{\text{good}})\right) \\ \nonumber
\,\text{for } i_{\text{bad}}\sim L=1,\, i_{\text{good}}\sim L=0 .
\end{align}
AUC has a direct operational interpretation: it is the probability that a randomly chosen bad item is ranked ahead of a randomly chosen good item. Thus, higher AUC corresponds to less manual effort to discover flawed items. Specifically, we simulate the workflow of a human reviewer who examines items in order of decreasing fit (i.e., starting with the most problematic items as identified by each metric).

Thus, the efficiency score is alternatively defined as:

$$\text{Efficiency} = \text{AUC} = \int_{0}^{1} \text{TPR}(\text{FPR}) \, d(\text{FPR})$$

where TPR (True Positive Rate) represents the proportion of truly problematic items identified, and FPR (False Positive Rate) represents the proportion of acceptable items incorrectly flagged. The estimation of the AUC was implemented with the \texttt{pROC} package \citep{robin_proc_2011}.

\paragraph{Sign convention.}
Many fit indices (including our signed isotonic coefficient) are ``higher is better.'' For AUC, we use a consistent convention that larger $S(i)$ means ``more suspicious'' by negating fit measures when needed.




\subsection{Competing methods}
\label{sec:baselines_exp}
We compare the signed isotonic $R^2$ coefficient against representative families of item-level diagnostics:

\begin{itemize}
    \item \textbf{Interitem association} (pairwise, aggregated to items): agreement-based statistics (e.g., SMC, $\kappa$), correlation-based statistics (e.g., $\phi$), and information-theoretic dependence (MI, symmetric uncertainty).
    \item \textbf{Item--rest statistics}: correlation or monotone dependence between item $i$ and the rest score $X_{-i}$.
    \item \textbf{Model-based parameters} (when feasible): IRT discrimination ($a$) and difficulty ($d$) parameters.
\end{itemize}

Our method is computed from interitem isotonic regressions (Sec.~\ref{sec:methods}); item-level scores are obtained by aggregating signed monotone $R^2_{i\rightarrow j}$ across $j\neq i$.


\begin{table*}[ht!]
\centering
\caption{Bad-item detection efficiency (AUC) for association-based baselines.
\textbf{Interitem} methods score an item by its mean pairwise association with all other items.
\textbf{Item-rest} methods score an item by its association with the rest-score $X_{-i}$.
Higher AUC indicates better prioritization of bad items for review.}
\label{tab:assoc_auc}
\begin{threeparttable}
\setlength{\tabcolsep}{6pt}
\renewcommand{\arraystretch}{1.15}

\begin{tabularx}{\textwidth}{@{} l X c c c c @{}}
\toprule
\textbf{Abbrev.} & \textbf{Statistic (pairwise)} & \textbf{HS Math} & \textbf{GSM8K} & \textbf{MMLU-5} & \textbf{Human} \\

\midrule
\multicolumn{5}{@{}l}{\textbf{\textit{Interitem comparisons}} (mean pairwise association for item $i$ with other items $j\neq i$)}\\
Acc & $\mathbb{P}(X_i=1, X_j=1)$ (probability both correct) & 0.585 & 0.781 & 0.770 & 0.667 \\
F1  &
$\displaystyle \frac{2\,\mathbb{P}(1\mid X_i=1)\,\mathbb{P}(1\mid X_j=1)}
{\mathbb{P}(1\mid X_i=1)+\mathbb{P}(1\mid X_j=1)}$
(cond. prob. both correct) & 0.627 & 0.821 & 0.791 & 0.683 \\
SMC & $\mathbb{P}(X_i = X_j)$ (probability of item agreement) & 0.885 & 0.841 & 0.827 & 0.733* \\
MI  & $H(X_i)+H(X_j)-H(X_i,X_j)$ (mutual dependence) & 0.814 & 0.863 & 0.763 & 0.817* \\
$\phi$ & $\mathrm{Corr}(X_i,X_j)$ (linear dependence) & 0.889 & \textit{0.872} & \textit{0.865} & 0.917* \\
$\rho_{tet}$ & $\mathrm{Corr}(\theta_i,\theta_j)$ (linear dependence among latent $\theta$s) & 0.765 & \textit{0.872} & \textit{0.866} & 0.896* \\
$\kappa$ &
$\displaystyle \frac{\mathbb{P}(X_i=X_j)-\mathbb{E}[\mathbb{P}(X_i=X_j)]}{1-\mathbb{E}[\mathbb{P}(X_i=X_j)]}$
(meas. error for same $\theta$) & 0.899 & 0.862 & \textbf{0.869} & 0.900* \\
$U_{sym}$ &
$\displaystyle \bar U_i=\frac{2}{n}\sum_{j\neq i}\frac{\mathrm{MI}(X_i;X_j)}{H(X_i)+H(X_j)}$
(symmetrized uncertainty) & 0.780 & \underline{\textbf{0.879}} & 0.681 & 0.883* \\
$\mathcal{M}_{\mathrm{iso}}$ &
Signed isotonic $R^2$ (prop. interitem signal explained) &
\underline{\textbf{0.908}} & \textbf{0.873} & \textit{0.861} & \underline{\textbf{0.983*}} \\
\midrule

\multicolumn{5}{@{}l}{\textbf{\textit{Item-group comparisons}} (association of item $i$ against all items/rest-score $X_{-i}$)}\\

$\text{MI}_{X,X(-i)}$ &
$H(X_i)+H(X_{-i})-H(X_i,X_{-i})$ (mutual information) & 0.576 & 0.553 & 0.597 & 0.590 \\
$\rho$ & $\mathrm{Corr}(X_i, X_{-i})$ (linear dependence) & 0.789 & \textit{0.870} & \underline{\textbf{0.871}} & 0.896* \\
$z$ &
$\displaystyle \sqrt{N-1}\;
\frac{\sum_{j\neq i}\mathrm{Cov}(X_i,X_j)}
{\sqrt{\sum_{j\neq i}\mathrm{Var}(X_i)\,\mathrm{Var}(X_j)}}$
(monotone dependence) & 0.788 & \textit{0.871} & \underline{\textbf{0.871}} & \textbf{0.933*} \\

${R}^2_{\mathrm{iso: X, i}}$  &
Signed isotonic $R^2$ (item signal explained by $X$) & \textbf{0.904} & \textbf{0.873} & 0.830 & 0.932* \\
\midrule

\multicolumn{5}{@{}l}{\textbf{\textit{Item-drop comparisons}} (difference in statistic upon removal of $i$)} \\

$\Delta \alpha$ &
Change in Reliability; $\alpha_X-\alpha_{X-i}$ & 0.805
  & \textbf{0.873} & 0.850  &  0.922* \\
  
$\Delta \bar\rho$ &
Change in Mean Correlation; $\bar\rho_X-\bar\rho_{X-i}$&
 0.788 & 0.870 & \underline{\textbf{0.871}}  &  0.895* \\

\midrule
\multicolumn{5}{@{}l}{\textbf{\textit{Item-as-parameter comparisons}} (Use of parameter value)}\\

$d_{\text{2PL}}$ &
2PL difficulty parameter; $\sigma(a_i(\theta-d_i))$ &
0.670 & 0.746 & 0.761 & 0.771 \\
$a_{\text{2PL}}$ &
2PL discrimination parameter; $\sigma(a_i(\theta-d_i))$ &
0.711 & \textbf{0.873} & 0.837 & 0.917* \\

\bottomrule
\end{tabularx}

\begin{tablenotes}[flushleft]\footnotesize
\item \underline{\textbf{Bold Underline}} indicates the best AUC for a given dataset; \textbf{Bold} indicates the second-best; \textit{italics} indicate others within $0.01$ from best. `*' signifies that for the first human dataset discussed in Sec. \ref{sec:humdata}, the one ``bad'' item was correctly sorted first; $H(\cdot)$ denotes Shannon entropy; $\mathrm{MI}(\cdot;\cdot)$ mutual information. $\sigma(\cdot)$ denotes sigmoid. ``tet.'' denotes tetrachoric correlation (latent $\theta$ correlation implied by dichotomous items).
\end{tablenotes}

\end{threeparttable}
\end{table*}

\section{Experiments}
\label{sec:experiments}

\subsection{Experiment 1: full-dataset AUC across analysis levels}
\label{sec:exp1}
The first experiment evaluates detection efficiency on the full datasets. For each method, we compute an item score on all five datasets and report the AUC (Table~\ref{tab:assoc_auc}). Our first experiment answers:
\emph{When the dataset is fixed, which statistics best prioritize globally bad items?}

\subsection{Experiment 2: subsampled benchmark stress test (AI)}
\label{sec:exp2}
AI benchmark regimes often face small-$n$/large-$p$ conditions: relatively few ``respondents'' (models, prompts, or runs) and many items. To probe robustness in this regime, we run $B=20$ subsampling trials for each AI benchmark. In each trial, we sample without replacement
\[
p=200\ \text{items} \quad\text{and}\quad n=50\ \text{respondents},
\]
compute a large battery of item statistics, and record AUC for each statistic on that subsample. We then aggregate performance across bootstraps by ranking statistics within each trial (by AUC) and averaging ranks; the comprehensive aggregated table is reported in Appendix Table~\ref{tab:efficiency_comparison}. This experiment evaluates the consistency of the many estimates in our statistical suite to generalize the statistical findings of the first experiment and reduce any sensitivity to particular combinations of items.

Thus, our second experiment answers:
\emph{Which methods remain reliable when both items and respondents are not fixed}

\subsection{Experiment 3: scalability across $n\times p$ regimes (interitem-only)}
\label{sec:exp3}
The third experiment isolates computationally light interitem methods to explore scaling behavior as the data aspect ratio varies. This setting is especially important for benchmarks that evolve over time (items added/removed) and for emerging evaluation settings where $n$ is limited.

We consider two AI benchmarks (GSM8K and MMLU) and generate $5600$ resamples total:
\begin{itemize}
    \item $p \in \{2^3,2^4,\dots,2^9\} = \{8,16,\dots,512\}$,
    \item $n$ varies over a grid of proportions of the original respondents, including deciles in $[0.4,1.0]$ and an extrapolated setting at $1.1$,
    \item for each $(n,p)$ configuration, we run $100$ resamples with replacement (subject to a minimum number of bad items).
\end{itemize}

For each resample, we compute AUC for each interitem statistic and then average AUC over the $100$ resamples per $(n,p)$. Finally, to summarize overall robustness across regimes, we apply a \emph{Borda count} aggregation: each $(n,p)$ configuration votes on a total ordering of methods by AUC, and we sum votes across configurations.

The third experiment answers:
\emph{Which interitem metrics remain strong across wide changes in sample size and test length?}

\subsection{Estimation details}
Detailed results are presented in the appendix, demonstrating the practical advantages of these novel scalability formulations for modern psychometric applications. Experiment 1 compares all the different computation costs: all interitem, except tetrachoric, relationships and nonparametric relationships were computed in \texttt{R} \citep{r_core_team_r_nodate} in 4.0, 8.1, and 22.4 seconds for HS math, MMLU, and GSM8K respectively; whereas tetrachoric alone took 8.2, 39.0, and 190.5 seconds using \texttt{psych} package \citep{revelle_psych_2024}; fitting 2PL models using \texttt{mirt} \citep{chalmers_mirt_2012} took 5.0, 24.5, and 15.4 seconds, respectively. Detailed results are presented in the appendix with all comparisons computed as measured under a bootstrapped Borda count regime for ranking across subsample sizes. AUC is computed using \texttt{pROC}~\citep{robin_proc_2011}. We drop degenerate cases where a method yields constant scores or where a resample contains only one label class. Missingness rates were recorded per metric per resample to distinguish statistical weakness from non-estimability. In the present study, missingness was negligible.

\section{Results}
\label{sec:results}

Results demonstrate substantial improvements in identification efficiency for both proposed metrics. Table \ref{tab:efficiency_comparison} presents AUC values across multiple datasets, with $\mathcal{M}_{\text{iso}}$ consistently ranking among the top-performing indices. 

\subsection{Experiment 1: signed isotonic \texorpdfstring{$R^2$}{R2} is consistently top-tier}
Table~\ref{tab:assoc_auc} reports AUC for representative interitem, item--rest, and model-based baselines across datasets. Three patterns are consistent.

\paragraph{(i) Interitem monotone variance explained is highly diagnostic.}
The signed isotonic $R^2$ achieves the best or near-best AUC across datasets, including the strongest performance on the human dataset (AUC $\approx 0.98$). This indicates that globally bad items are precisely those that fail to participate in the monotone dependency structure shared by most items.

\paragraph{(ii) Nonlinear monotone signal matters beyond linear correlation.}
Linear association ($\phi$) and agreement-based measures (SMC, $\kappa$) are competitive on some datasets, but the signed isotonic $R^2$ improves or matches them without assuming linearity and while preserving the direction of contribution. This is particularly important when misfit produces nonlinearity (e.g., ambiguity affecting mid-ability respondents disproportionately).

\paragraph{(iii) Item--rest summaries can underperform interitem structure.}
Item--rest mutual information is substantially weaker than the best interitem methods in these datasets. Aggregating through the rest score $X_{-i}$ can wash out pairwise violations, especially when a small number of bad items is diluted by many good ones.

\subsection{Experiment 2: robustness under subsampling and broad method comparison}
Appendix Table~\ref{tab:efficiency_comparison} reports results from a comprehensive battery of indices (CTT, IRT, scalability, PCA/omega, anomaly metrics) under $B=20$ subsamples per AI benchmark. Signed isotonic $R^2$ ranks among the most efficient methods by average rank and percentile-AUC summaries, outperforming many widely used diagnostics.

Two observations are particularly practically relevant:

\paragraph{Stability under benchmark-sized subsamples.}
When restricted to only $n=50$ respondents and $p=200$ items, many model-based procedures can become unstable, slow, or sensitive to estimation choices. The signed isotonic $R^2$ remains competitive because it depends only on bivariate monotone fits and aggregation.

\paragraph{Comparable accuracy with substantially lighter computation.}
On the full datasets, computing the full set of interitem statistics (including isotonic regression) required seconds, whereas tetrachoric correlations and some IRT fits required substantially longer. This supports the use of signed isotonic $R^2$ as an early-stage screening tool: it produces high-quality review queues cheaply.

\subsection{Experiment 3: best overall scaling robustness across \texorpdfstring{$n\times p$}{n x p} regimes}
Across $5600$ resamples spanning $p$ up to $512$ and varying $n$, Borda aggregation yields a clear ordering: the signed isotonic $R^2$ is the most robust interitem method overall, followed by the signed adjusted isotonic $R^2$, then correlation-based ($\phi$/MCC) and classic agreement-based metrics.

This result is consistent with the interpretation that signed isotonic $R^2$ captures a fundamental regularity of unidimensional measurement--monotone coherence--while remaining insensitive to functional-form misspecification and changes in aspect ratio.

Proposition~\ref{prop:max_monotone_r2} shows that the signed isotonic $R^2$ is not merely another association coefficient: it is an \emph{extremal} measure of monotone predictability. When most items share a latent ordering, a good item should be predictably monotone from many others; globally bad items systematically reduce this maximal monotone predictability.

\section{Discussion}
\label{sec:discussion}

\subsection{What the experiments collectively show}
Across full datasets (Exp.~1), stress-tested subsamples (Exp.~2), and wide aspect-ratio regimes (Exp.~3), signed isotonic $R^2$ is consistently among the best detectors of globally bad items. These findings support two claims:

\begin{enumerate}
    \item \textbf{Measurement coherence is fundamentally monotone.} The strongest signal distinguishing good from bad items is whether an item participates in the expected monotone dependency structure induced by a dominant latent trait.
    \item \textbf{Maximal monotone explainability is a practical screening principle.} Quantifying ``how much of an item can be monotone-explained by other items'' yields review queues that are efficient across diverse item pathologies.
\end{enumerate}

\subsection{Why signed isotonic \texorpdfstring{$R^2$}{R2} detects global badness}
Bad items can fail in different ways (bad key, grading, ambiguity, construct drift), but they share a common footprint: they \emph{break monotone coherence} with the rest of the instrument.

\paragraph{Directionality matters.}
Miskeyed or systematically inverted grading induces \emph{negative} association with many items. A signed statistic can surface these inversions directly, whereas unsigned dependence measures may treat inversions as ``strong signal'' and mis-rank the item. This is one reason the signed isotonic family tends to outperform the unsigned isotonic $R^2$ (which, for dichotomous items, collapses toward a squared correlation-style quantity).

\paragraph{Nonlinearity matters.}
Ambiguity and construct drift can produce non-monotone or saturating effects: an item may behave normally for low-ability respondents but become noisy for high-ability respondents (or vice versa). Linear correlation averages these regimes; isotonic regression explicitly extracts the strongest monotone component and penalizes deviations through reduced $R^2$.

\subsection{Implications for AI benchmarks}
AI benchmarks lack the institutionalized validation pipelines typical of human testing. Our results suggest a lightweight, model-agnostic workflow:

\begin{enumerate}
    \item Compute signed isotonic $R^2$ item scores from an evaluation matrix (models$\times$items or runs$\times$items).
    \item Review the top-$k$ most suspicious items.
    \item Fix or remove problematic items; re-score models; repeat.
\end{enumerate}

Because the method is bivariate and nonparametric, it can be applied even when item formats vary (binary/ordinal/continuous) and when the ``respondents'' are heterogeneous systems rather than humans.

\subsection{How to read Appendix Table~\ref{tab:efficiency_comparison}}
The appendix table is intentionally comprehensive: it reports (i) a large suite of indices spanning interitem association, Mokken scalability, item-drop reliability, IRT item fit, and dimensionality proxies, and (ii) aggregated performance summaries across subsamples. Rather than serving as a central narrative element, it functions as a \emph{robustness audit}: the signed isotonic $R^2$ remains highly ranked even against dozens of alternative diagnostics and under repeated perturbations of the dataset.

In the main text we therefore emphasize:
(i) the head-to-head comparisons most interpretable to a broad audience (Table~\ref{tab:assoc_auc}),
and (ii) the scaling regime where AI benchmark practice is most challenged (Exp.~2--3).

\subsection{Conclusion}
Across both human assessments and AI benchmarks, signed isotonic $R^2$ provides a simple rule with a strong empirical and mathematical basis:

\begin{quote}
\emph{Items that cannot be well explained by monotone functions of other items are the ones most likely to be globally bad.}
\end{quote}

This principle yields review queues that are more efficient, more scalable, and less assumption-laden than many traditional alternatives, making it well-suited for modern large-scale evaluation regimes.

\subsection{Limitations and scope}
\paragraph{Global detection, not diagnosis.}
High suspicion scores indicate that an item fails to cohere with the rest of the instrument, but they do not identify the cause (miskey vs.\ ambiguity vs.\ construct drift). In practice, the score is a prioritization tool for human review.

\paragraph{Not a DIF tool.}
Because we do not condition on group membership, this method does not detect group-conditional misfit. Combining signed isotonic $R^2$ with stratified analyses is a promising extension.

\paragraph{Dependence on a dominant monotone structure.}
If the instrument is strongly multidimensional or intentionally non-monotone, interitem monotone coherence is not the correct baseline and may over-flag items. In such cases, applying the method within clusters (e.g., topics/subscales) or after dimensionality screening is recommended.

\section*{Acknowledgments}
We'd like to thank Lijin Zhang, Sanmi Koyejo, Sang Truong, Yuheng Tu, Elizabeth Childs, Yunsung Kim, Anka Reuel, Hansol Lee, and Jason Cho for their time, input, and support.

\bibliography{references}
\bibliographystyle{icml2026}

\newpage
\appendix
\onecolumn
\section{Appendix}

\begingroup\fontsize{7}{9}\selectfont\label{tab:efficiency_comparison}
\begin{longtable}{llrrrr}
\toprule
Metric & Metric Type & Ave. Rank & $R_{1.0}\text{MMLU5}$ & $R_{1.0}\text{GSM8K}$ & $R_{1.0}\text{HSMath}$ \\
$\mathcal{M}_{iso}$& Isotonic& 12.67& 0.91 & 0.83 & 0.85\\
c75 & Inter-Item Correlations (tet), 75-quantile& 14.00& 0.91 & 0.83 & 0.84\\
$a_{2PL}$& IRT: Discrimination & 19.33 & 0.91 & 0.83 & 0.83\\
$g_{(5)}$ & Omega: General Factor& 25.00 & 0.90 & 0.82 & 0.81\\
$g_{(3)}$ & Omega: General Factor& 25.00 & 0.88 & 0.81 & 0.90\\
$h^2_{(3)}$ & Omega: Variance & 25.50 & 0.73 & 0.82 & 0.95\\
\addlinespace
$u^2_{(3)}$ & PCA: Variance & 26.83 & 0.73 & 0.82 & 0.95\\
$h^2_{(5)}$ & Omega: Variance & 26.83 & 0.73 & 0.82 & 0.96\\
min($\rho_{i(tet)}$) & Inter-Item Correlations (tet) & 27.33 & 0.90 & 0.81 & 0.86\\
\addlinespace
$u^2_{(5)}$ & PCA: Variance & 28.50 & 0.72 & 0.82 & 0.96\\
$\rho_\tau$ mean& Inter-Item Correlations (tetrachoric)& 28.83 & 0.91 & 0.83 & 0.77\\
$M_{\tau,j}$& Kendall's $\tau$& 30.00 & 0.91 & 0.83 & 0.77\\
$a_{3PL}$& IRT: Discrimination & 31.83 & 0.84 & 0.79 & 0.89\\
\addlinespace
$h^2_{(15)}$ & Omega: Variance & 31.83 & 0.67 & 0.82 & 0.96\\
$\Delta\alpha$ & CTT: Reliability & 32.17 & 0.86 & 0.83 & 0.78\\
Zi & Scalability & 34.00 & 0.90 & 0.83 & 0.77\\
$\Delta r$ & other & 34.00 & 0.90 & 0.83 & 0.77\\
\addlinespace
$u^2_{(15)}$ & PCA: Variance & 34.33 & 0.67 & 0.81 & 0.96\\
cmedse & Inter-Item Correlations (tet) & 35.00 & 0.69 & 0.76 & 0.97\\
Hi (Loevinger) & Scalability & 35.33 & 0.91 & 0.79 & 0.79\\
$g_{(15)}$ & other & 36.83 & 0.91 & 0.83 & 0.73\\
SE($\rho$) & Inter-Item Correlations (tet) & 37.83 & 0.70 & 0.78 & 0.94\\
\addlinespace
Var($\rho$) & Inter-Item Correlations (tet) & 37.83 & 0.70 & 0.78 & 0.94\\
median($\rho$) & Inter-Item Correlations (tet) & 41.17 & 0.90 & 0.82 & 0.72\\
dcor\_drop & Rest Distance Correlation & 42.17 & 0.77 & 0.82 & 0.77\\
p2\_5 & Omega: Variance & 48.17 & 0.73 & 0.76 & 0.81\\
\addlinespace
p2\_3 & Omega: Variance & 48.67 & 0.62 & 0.69 & 0.97\\
P & manifest & 50.67 & 0.91 & 0.77 & 0.66\\
cneg & Inter-Item Correlations (tet) & 50.67 & 0.89 & 0.80 & 0.67\\
corr\_drop & Rest Correlation & 50.67 & 0.86 & 0.80 & 0.70\\
c25 & Inter-Item Correlations (tet) & 51.33 & 0.89 & 0.82 & 0.60\\
\addlinespace
com\_15 & Omega: Variance & 51.83 & 0.58 & 0.72 & 0.96\\
tc5\_15 & PCA: Variance & 54.00 & 0.63 & 0.65 & 0.92\\
p2\_15 & Omega: Variance & 56.17 & 0.75 & 0.79 & 0.73\\
c5 & Inter-Item Correlations (tet) & 58.83 & 0.80 & 0.81 & 0.63\\
g2\_2 & Omega: Variance & 59.00 & 0.57 & 0.68 & 0.95\\
\addlinespace
$\Sigma \text{Viol.}$ & Scalability (Mokken) & 59.50 & 0.83 & 0.81 & 0.61\\
f3\_5 & Omega: Variance & 61.67 & 0.64 & 0.64 & 0.83\\
tc13\_15 & PCA: Variance & 61.67 & 0.64 & 0.61 & 0.91\\
c10 & Inter-Item Correlations (tet) & 62.33 & 0.86 & 0.82 & 0.51\\
\addlinespace
tc11\_15 & PCA: Variance & 63.33 & 0.71 & 0.47 & 0.96\\
f13\_15 & Omega: Variance & 64.17 & 0.67 & 0.59 & 0.87\\
tc4\_5 & PCA: Variance & 64.33 & 0.64 & 0.67 & 0.80\\
raw\_alpha & Item-drop & 64.50 & NA & NA & 0.78\\
$N_{\text{viol.}}$ & Scalability & 65.17 & 0.85 & 0.81 & 0.51\\
\addlinespace
f15\_15 & Omega: Variance & 65.67 & 0.68 & 0.53 & 0.90\\
crit & Inter-Item Correlations (tet) & 66.33 & 0.86 & 0.77 & 0.53\\
tc2\_15 & PCA: Variance & 66.67 & 0.67 & 0.58 & 0.85\\
adj\_depth & Anomaly: Item Isolation & 66.67 & 0.46 & 0.82 & 0.81\\
com\_3 & Omega: Variance & 67.33 & 0.60 & 0.68 & 0.82\\
\addlinespace
f1\_5 & Omega: Variance & 67.50 & 0.81 & 0.62 & 0.69\\
Max Viol. & Scalability (Mokken) & 67.67 & 0.72 & 0.74 & 0.65\\
tc7\_15 & PCA: Variance & 68.00 & 0.60 & 0.63 & 0.86\\
tc2\_3 & PCA: Variance & 68.33 & 0.74 & 0.62 & 0.76\\
x2\_2 & IRT: Item Fit & 68.50 & 0.54 & 0.65 & 0.88\\
\addlinespace
f2\_3 & Omega: Variance & 68.67 & 0.74 & 0.62 & 0.76\\
f11\_15 & Omega: Variance & 68.67 & 0.61 & 0.60 & 0.91\\
adj\_density & Anomaly: Item Isolation & 69.00 & 0.47 & 0.82 & 0.79\\
sum\_number\_ac & Scalability (Mokken) & 69.33 & 0.82 & 0.75 & 0.54\\
f12\_15 & Omega: Variance & 69.33 & 0.67 & 0.54 & 0.87\\
\addlinespace
number\_ac & Scalability & 69.33 & 0.66 & 0.74 & 0.73\\
number\_vi\_number\_ac & Scalability & 69.83 & 0.81 & 0.72 & 0.58\\
f9\_15 & Omega: Variance & 69.83 & 0.55 & 0.60 & 0.95\\
f2\_5 & Omega: Variance & 70.67 & 0.74 & 0.67 & 0.63\\
tc1\_5 & PCA: Variance & 71.17 & 0.81 & 0.64 & 0.63\\
\addlinespace
tc2\_5 & PCA: Variance & 72.00 & 0.74 & 0.62 & 0.68\\
tc1\_3 & PCA: Variance & 72.33 & 0.82 & 0.68 & 0.56\\
tc1\_15 & PCA: Variance & 73.67 & 0.60 & 0.62 & 0.80\\
f1\_3 & Omega: Variance & 75.00 & 0.82 & 0.64 & 0.56\\
\addlinespace
tc5\_5 & PCA: Variance & 75.33 & 0.62 & 0.60 & 0.81\\
f2\_15 & Omega: Variance & 75.67 & 0.65 & 0.56 & 0.80\\
average\_r & other & 76.00 & NA & NA & 0.77\\
g6\_smc & CTT: Reliability & 76.00 & NA & NA & 0.77\\
r\_cor & Item-total Correlation & 76.00 & NA & NA & 0.77\\
\addlinespace
r\_drop\_alpha & Item-drop & 76.00 & NA & NA & 0.77\\
raw\_r & Item-drop & 76.00 & NA & NA & 0.77\\
SNR & Alpha & 76.00 & NA & NA & 0.77\\
std\_alpha & Item-drop & 76.00 & NA & NA & 0.77\\
std\_r & Item-drop & 76.00 & NA & NA & 0.77\\
\addlinespace
f5\_15 & Omega: Variance & 76.67 & 0.63 & 0.64 & 0.74\\
rmsea\_x2\_2 & Fit Deviation & 76.83 & 0.50 & 0.61 & 0.88\\
f4\_5 & Omega: Variance & 77.67 & 0.62 & 0.60 & 0.80\\
complexity\_5 & PCA: Variance & 77.83 & 0.62 & 0.51 & 0.87\\
f8\_15 & Omega: Variance & 79.00 & 0.56 & 0.60 & 0.86\\
\addlinespace
outfit\_3 & IRT: Item Fit & 80.33 & 0.56 & 0.74 & 0.73\\
f3\_15 & Omega: Variance & 81.17 & 0.67 & 0.61 & 0.69\\
f1\_15 & Omega: Variance & 82.17 & 0.63 & 0.59 & 0.78\\
rmsea\_g2\_2 & Fit Deviation & 82.33 & 0.57 & 0.53 & 0.87\\
tc3\_5 & PCA: Variance & 82.33 & 0.49 & 0.62 & 0.83\\
\addlinespace
c1 & Inter-Item Correlations (tet) & 83.33 & 0.61 & 0.80 & 0.57\\
tc10\_15 & PCA: Variance & 83.67 & 0.62 & 0.53 & 0.80\\
tc12\_15 & PCA: Variance & 84.33 & 0.54 & 0.59 & 0.84\\
com\_5 & Omega: Variance & 85.17 & 0.58 & 0.70 & 0.66\\
complexity\_15 & PCA: Variance & 86.00 & 0.58 & 0.51 & 0.85\\
\addlinespace
tc6\_15 & PCA: Variance & 86.17 & 0.62 & 0.48 & 0.81\\
cmin & Inter-Item Correlations (tet) & 86.67 & 0.54 & 0.50 & 0.90\\
f5\_5 & Omega: Variance & 87.00 & 0.47 & 0.62 & 0.80\\
z\_infit\_3 & IRT: Item Fit & 87.67 & 0.78 & 0.54 & 0.60\\
f14\_15 & Omega: Variance & 88.33 & 0.66 & 0.64 & 0.56\\
\addlinespace
tc15\_15 & PCA: Variance & 88.33 & 0.63 & 0.63 & 0.61\\
f3\_3 & Omega: Variance & 88.50 & 0.47 & 0.62 & 0.79\\
rmsea\_g2\_3 & Fit Deviation & 88.67 & 0.58 & 0.55 & 0.79\\
g2\_3 & Omega: Variance & 89.00 & 0.55 & 0.71 & 0.64\\
tc3\_3 & PCA: Variance & 91.50 & 0.47 & 0.61 & 0.79\\
\addlinespace
tc3\_15 & PCA: Variance & 92.50 & 0.52 & 0.59 & 0.78\\
tc14\_15 & PCA: Variance & 94.00 & 0.69 & 0.55 & 0.59\\
z\_outfit\_3 & IRT: Item Fit & 94.17 & 0.59 & 0.62 & 0.64\\
f10\_15 & Omega: Variance & 94.50 & 0.46 & 0.61 & 0.77\\
guess\_3 & IRT: Discrimination & 95.17 & 0.68 & 0.46 & 0.66\\
\addlinespace
outfit\_2 & IRT: Item Fit & 95.17 & 0.53 & 0.70 & 0.61\\
tc9\_15 & PCA: Variance & 95.50 & 0.66 & 0.52 & 0.64\\
infit\_3 & IRT: Item Fit & 95.67 & 0.78 & 0.52 & 0.53\\
x2\_3 & IRT: Item Fit & 95.67 & 0.50 & 0.69 & 0.63\\
tc8\_15 & PCA: Variance & 99.67 & 0.63 & 0.60 & 0.54\\
\addlinespace
rmsea\_x2\_3 & Fit Deviation & 100.33 & 0.49 & 0.64 & 0.63\\
complexity\_3 & PCA: Variance & 100.83 & 0.63 & 0.58 & 0.58\\
z\_infit\_2 & IRT: Item Fit & 103.00 & 0.64 & 0.50 & 0.61\\
f7\_15 & Omega: Variance & 104.17 & 0.60 & 0.61 & 0.51\\
\addlinespace
f4\_15 & Omega: Variance & 106.00 & 0.51 & 0.52 & 0.75\\
infit\_2 & IRT: Item Fit & 106.17 & 0.60 & 0.52 & 0.63\\
f6\_15 & Omega: Variance & 107.50 & 0.63 & 0.46 & 0.63\\
z\_outfit\_2 & IRT: Item Fit & 110.67 & 0.55 & 0.59 & 0.56\\
number\_zsig & Scalability & 111.33 & 0.62 & 0.56 & 0.42\\
\addlinespace
tc4\_15 & PCA: Variance & 111.33 & 0.56 & 0.56 & 0.56\\
cmax & Inter-Item Correlations (tet) & 112.00 & 0.51 & 0.53 & 0.63\\
MI\_drop & Rest Mutual Information & 118.00 & 0.56 & 0.52 & 0.52\\
msa\_15 & PCA: Variance & 127.50 & 0.50 & 0.50 & 0.50\\
msa\_3 & PCA: Variance & 127.50 & 0.50 & 0.50 & 0.50\\
\addlinespace
msa\_5 & PCA: Variance & 127.50 & 0.50 & 0.50 & 0.50\\
\bottomrule
\caption{The first numeric column represents the average Borda count rank. The subsequent columns represent the average percentile rank for AUC. Estimation of the AUC was done with the \texttt{pROC} package \citep{robin_proc_2011}.}
\end{longtable}
\endgroup{}

\begin{figure}
    \centering
    \includegraphics[width=1\linewidth]{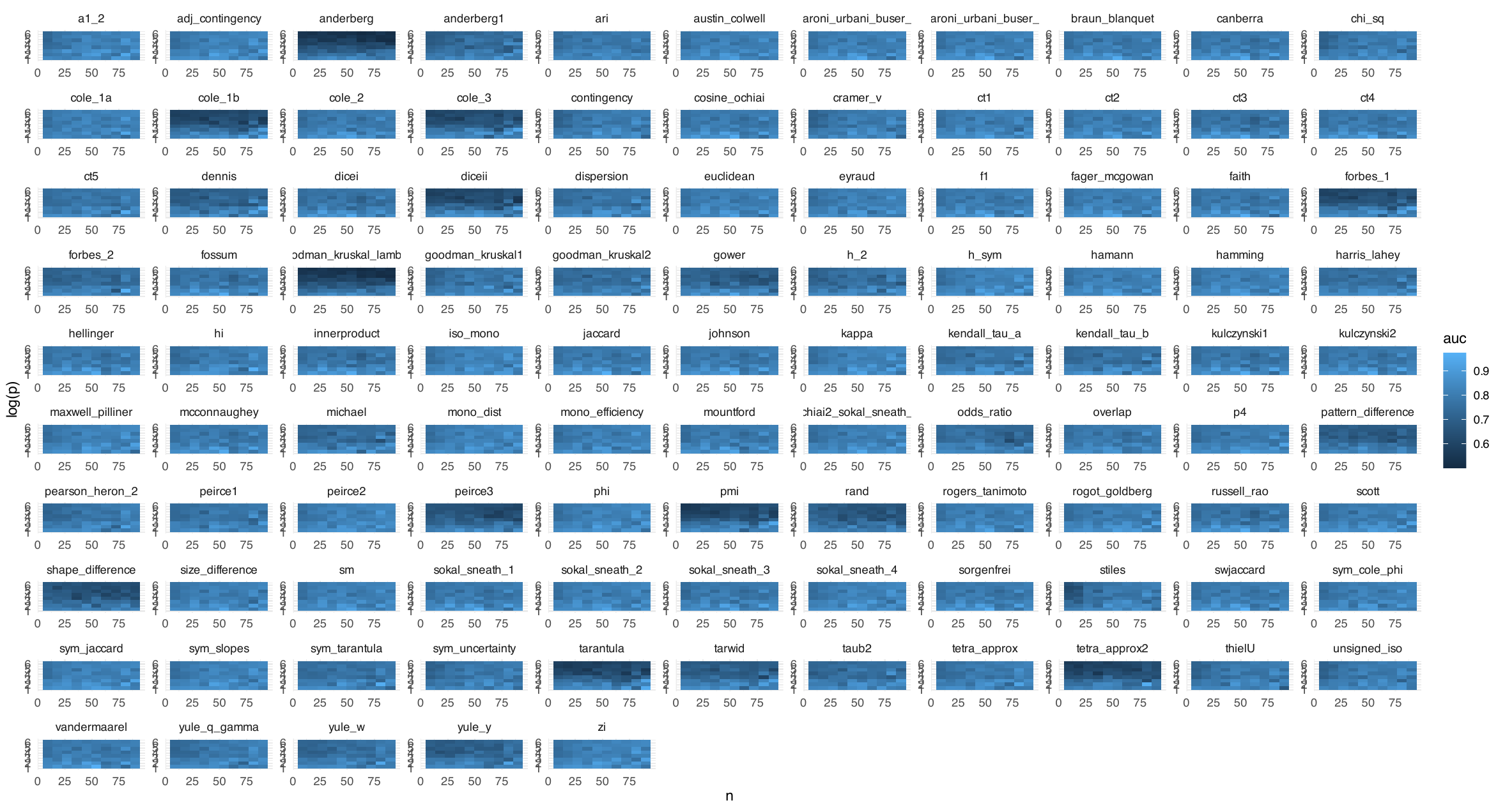}
    \caption{Instrument Composition Permutation Bootstrap Results (n x p)}
    \label{fig:permboot}
\end{figure}

\section{Full List of Binary Associations}\label{apx:dichbinlist}

\input{tables/binary_associations}

\input{apx/measure_theory}

\end{document}

%% file: tables/binary_associations.tex



\begin{longtable}{>{\raggedright}p{3cm} >{\centering\arraybackslash}p{4.5cm} >{\raggedright\arraybackslash}p{6cm} >{\centering\arraybackslash}p{0.7cm}}
\caption{Intersection and Agreement-Based Similarity Measures}
\label{tab:agreement_measures} \\
\toprule
\textbf{Measure (Alias)} & \textbf{Expression} & \textbf{Description \& Rationale} & \textbf{Ref.} \\
\midrule
\endfirsthead

\multicolumn{4}{c}%
{{\tablename\ \thetable{} -- continued from previous page}} \\
\toprule
\textbf{Measure (Alias)} & \textbf{Expression} & \textbf{Description \& Rationale} & \textbf{Ref.} \\
\midrule
\endhead

\bottomrule
\multicolumn{4}{r}{{Continued on next page}} \\
\endfoot

\endlastfoot

\multicolumn{4}{l}{\textit{\textbf{Group 1: Intersection-Based (Focus on Joint Presence, `a')}}} \\ \addlinespace[0.3em]

Jaccard Index & $\displaystyle \frac{a}{a+b+c}$ & Ratio of the intersection to the union of positive responses. Ignores joint absences (\textit{d}), making it suitable when co-presence is the primary signal of similarity. & \cite{} \\ \addlinespace[0.5em]

Dice-Sørensen Coefficient (F1 Score) & $\displaystyle \frac{2a}{2a+b+c}$ & The harmonic mean of precision and recall. It is monotonic with Jaccard but gives greater weight to the intersection term (\textit{a}). & \cite{} \\ \addlinespace[0.5em]

Overlap Coefficient (Szymkiewicz-Simpson) & $\displaystyle \frac{a}{\min(a+b, a+c)}$ & An asymmetric measure of inclusion; represents the proportion of the smaller set that is contained within the larger set. Reaches 1 if one item's endorsers are a subset of the other's. & \cite{} \\ \addlinespace[0.5em]

Cosine Similarity (Ochiai, Fowlkes-Mallows) & $\displaystyle \frac{a}{\sqrt{(a+b)(a+c)}}$ & Geometric mean normalization. Interprets response vectors in a high-dimensional space and measures the cosine of the angle between them. Insensitive to vector magnitude. & \cite{} \\ \addlinespace[0.5em]

Kulczyński 1 & $\displaystyle \frac{a}{b+c}$ & Ratio of joint presences to disagreements. Highly sensitive to low frequencies of disagreement, diverging as \textit{b+c} approaches zero. & \cite{} \\ \addlinespace[0.5em]

Forbes I & $\displaystyle \frac{n \cdot a}{(a+b)(a+c)}$ & Ratio of the observed frequency of joint presence (\textit{a}) to its expected frequency under statistical independence. A value > 1 indicates positive association. & \cite{} \\ \addlinespace[0.5em]

\midrule
\multicolumn{4}{l}{\textit{\textbf{Group 2: Agreement-Based (Focus on Agreement, `a+d')}}} \\ \addlinespace[0.3em]

Simple Matching (Sokal-Michener) & $\displaystyle \frac{a+d}{n}$ & The most straightforward measure of agreement; the proportion of total cases where both items yield the same outcome. Treats joint presence and absence as equally informative. & \cite{} \\ \addlinespace[0.5em]

Rogers-Tanimoto & $\displaystyle \frac{a+d}{a+d+2(b+c)}$ & A variant of Simple Matching that penalizes disagreements by giving them double weight in the denominator. This results in lower similarity values compared to Simple Matching. & \cite{} \\ \addlinespace[0.5em]

Hamann & $\displaystyle \frac{(a+d) - (b+c)}{n}$ & Proportion of agreement minus the proportion of disagreement. Ranges from -1 (perfect disagreement) to +1 (perfect agreement). A signed measure sensitive to the balance of concordant and discordant pairs. & \cite{} \\ \addlinespace[0.5em]

Sokal-Sneath 2 (Gower-Legendre) & $\displaystyle \frac{2(a+d)}{2(a+d)+b+c}$ & A variant of the Dice-Sørensen coefficient that is applied to both agreement states (\textit{a} and \textit{d}). & \cite{} \\ \addlinespace[0.5em]

Faith & $\displaystyle \frac{a+0.5d}{n}$ & Asymmetrically weights agreement, considering a joint presence (\textit{a}) twice as informative as a joint absence (\textit{d}). Useful when positive matches are considered stronger evidence of similarity. & \cite{} \\
\bottomrule
\end{longtable}

\begin{longtable}{>{\raggedright}p{3cm} >{\centering\arraybackslash}p{4.5cm} >{\raggedright\arraybackslash}p{6cm} >{\centering\arraybackslash}p{0.7cm}}
\caption{Covariance, Correlation, and Information-Theoretic Measures}
\label{tab:correlation_measures} \\
\toprule
\textbf{Measure (Alias)} & \textbf{Expression} & \textbf{Description \& Rationale} & \textbf{Ref.} \\
\midrule
\endfirsthead

\multicolumn{4}{c}%
{{\tablename\ \thetable{} -- continued from previous page}} \\
\toprule
\textbf{Measure (Alias)} & \textbf{Expression} & \textbf{Description \& Rationale} & \textbf{Ref.} \\
\midrule
\endhead

\bottomrule
\multicolumn{4}{r}{{Continued on next page}} \\
\endfoot

\endlastfoot

\multicolumn{4}{l}{\textit{\textbf{Group 3: Covariance and Correlation-Based (Focus on `ad-bc')}}} \\ \addlinespace[0.3em]

Phi ($\phi$) Coefficient (Matthews Correlation) & $\displaystyle \frac{ad-bc}{\sqrt{(a+b)(c+d)(a+c)(b+d)}}$ & The Pearson product-moment correlation for two dichotomous variables. It is a chance-corrected measure of association, sensitive to both marginal distributions and the covariance term. & \cite{} \\ \addlinespace[0.5em]

Yule's Q ($\gamma$) & $\displaystyle \frac{ad-bc}{ad+bc}$ & A measure of ordinal association that is independent of the marginal distributions. It represents the probability of concordance minus the probability of discordance, given that no ties exist. Reaches $\pm 1$ if any cell is zero. & \cite{} \\ \addlinespace[0.5em]

Yule's W & $\displaystyle \frac{\sqrt{ad}-\sqrt{bc}}{\sqrt{ad}+\sqrt{bc}}$ & The "coefficient of colligation." It is more conservative than Yule's Q and is not independent of the marginals. It relates to the average of conditional probabilities. & \cite{} \\ \addlinespace[0.5em]

Odds Ratio (OR) & $\displaystyle \frac{ad}{bc}$ & The ratio of the odds of a positive outcome on Item \textit{j} given a positive outcome on Item \textit{i}, to the odds of a positive outcome on \textit{j} given a negative outcome on \textit{i}. Widely used in epidemiology; not bounded above. & \cite{} \\ \addlinespace[0.5em]

Cohen's Kappa ($\kappa$) & $\displaystyle \frac{p_o - p_e}{1 - p_e}$ \newline \small where $p_o=\frac{a+d}{n}$ and \newline $p_e=\frac{(a+b)(a+c)+(c+d)(b+d)}{n^2}$ & A measure of inter-rater reliability that corrects the observed proportion of agreement ($p_o$) for the agreement expected by chance ($p_e$). Robust against unequal marginal probabilities. & \cite{} \\ \addlinespace[0.5em]

Dispersion & $\displaystyle \frac{ad-bc}{n^2}$ & The covariance between the two binary items. It is the fundamental building block of many correlation coefficients but is sensitive to the variance of the items. & \cite{} \\ \addlinespace[0.5em]

\midrule
\multicolumn{4}{l}{\textit{\textbf{Group 4: Information-Theoretic}}} \\ \addlinespace[0.3em]

Pointwise Mutual Information (PMI) & $\displaystyle \log_2\left(\frac{P(i,j)}{P(i)P(j)}\right) = \log_2\left(\frac{a \cdot n}{(a+c)(a+b)}\right)$ & Quantifies the discrepancy between the probability of the items' co-occurrence given their joint distribution versus their individual distributions (assuming independence). & \cite{} \\ \addlinespace[0.5em]

Symmetric Uncertainty & $\displaystyle \frac{2 \cdot I(i;j)}{H(i)+H(j)}$ & A normalized variant of mutual information, $I(i;j)$. It corrects for the bias of mutual information toward variables with more states (though not an issue for binary data) and produces a value in [0, 1]. & \cite{} \\
\bottomrule
\end{longtable}

\begin{longtable}{>{\raggedright}p{3cm} >{\centering\arraybackslash}p{4.5cm} >{\raggedright\arraybackslash}p{6cm} >{\centering\arraybackslash}p{0.7cm}}
\caption{Psychometric and Monotonicity-Based Measures}
\label{tab:psychometric_measures} \\
\toprule
\textbf{Measure (Alias)} & \textbf{Expression} & \textbf{Description \& Rationale} & \textbf{Ref.} \\
\midrule
\endfirsthead

\multicolumn{4}{c}%
{{\tablename\ \thetable{} -- continued from previous page}} \\
\toprule
\textbf{Measure (Alias)} & \textbf{Expression} & \textbf{Description \& Rationale} & \textbf{Ref.} \\
\midrule
\endhead

\bottomrule
\endfoot

Loevinger's H (Pairwise, $H_{ij}$) & $\displaystyle \frac{\text{Cov}(i,j)}{\text{Cov}_{\text{max}}(i,j)}$ \newline \small For $\text{Cov}>0$: $\displaystyle \frac{ad-bc}{\min\left[(a+c)(b+d), (a+b)(c+d)\right]}$ & A coefficient of homogeneity. It normalizes the covariance by the maximum possible covariance given the items' marginal frequencies. This assesses how well the items conform to a deterministic Guttman-type relationship. A cornerstone of Mokken Scale Analysis. & \cite{} \\ \addlinespace[0.5em]

Standardized Z ($Z_i$) & \small Conceptually: $\displaystyle \frac{H_i - E[H_i]}{\sqrt{\text{Var}(H_i)}}$ & A test statistic for the significance of an item's scalability ($H_i$, its average H-value with all other items). It assesses whether an item's monotonicity with the underlying scale is statistically significant under the null model of non-monotonicity. Not directly computed from a single pair. & \cite{} \\ \addlinespace[0.5em]

2PL Discrimination ($a_i$) & \small Estimated via Marginal Maximum Likelihood (MML) & A parameter from a latent trait (Item Response Theory) model. It represents the slope of the Item Characteristic Curve at the point of inflection ($\theta = b_i$). It is not expressible in contingency table terms, as it is estimated from the full response matrix under the assumption of a continuous latent trait. It is analogous to a factor loading. & \cite{} \\ \addlinespace[0.5em]

Yule's Q ($\gamma$) & $\displaystyle \frac{ad-bc}{ad+bc}$ & Included again for comparison. In psychometrics, its property of being independent of marginals makes it a pure measure of ordinal association, but this can be a drawback, as it may inflate association for items with extreme difficulties. & \cite{} \\ \addlinespace[0.5em]

Phi ($\phi$) Coefficient & $\displaystyle \frac{ad-bc}{\sqrt{(a+b)(c+d)(a+c)(b+d)}}$ & Included again for comparison. In contrast to Q, Phi is sensitive to item difficulty differences. Two items with high Phi must have similar difficulties, making it a measure of both association and difficulty-matching. Its square is the proportion of variance explained. & \cite{} \\ \addlinespace[0.5em]

\bottomrule
\end{longtable}

\begin{table}[htbp]
\centering
\caption{Binary Item Similarity Measures: Mathematical Formulations and Characteristics}
\label{tab:similarity_measures}
\resizebox{\textwidth}{!}{%
\begin{tabular}{>{\raggedright}p{2.8cm} >{\raggedright}p{2.2cm} >{\centering}p{4.5cm} >{\raggedright}p{5.8cm} p{1.2cm}}
\toprule
\textbf{Measure} & \textbf{Aliases} & \textbf{Formula} & \textbf{Description \& Key Features} & \textbf{Refs} \\
\midrule
\multicolumn{5}{l}{\textit{\textbf{Intersection-Based Measures}}} \\
\midrule
Jaccard Index & Tanimoto & $\frac{a}{a+b+c}$ & Intersection over union; satisfies triangle inequality; standard for binary similarity & \\
\addlinespace[0.5ex]
Dice Coefficient & Sørensen, F1-score & $\frac{2a}{2a+b+c}$ & Harmonic mean of precision/recall; emphasizes joint occurrences & \\
\addlinespace[0.5ex]
Overlap Coefficient & Simpson & $\frac{a}{\min(a+b, a+c)}$ & Similarity relative to smaller set; asymmetric upper bound & \\
\addlinespace[0.5ex]
Cosine Similarity & Ochiai, Driver-Kroeber & $\frac{a}{\sqrt{(a+b)(a+c)}}$ & Geometric angle measure; normalized dot product for binary vectors & \\
\addlinespace[0.5ex]
Kulczyński-2 & & $\frac{1}{2}\left(\frac{a}{a+b} + \frac{a}{a+c}\right)$ & Arithmetic mean of conditional probabilities; symmetric precision measure & \\
\midrule
\multicolumn{5}{l}{\textit{\textbf{Agreement-Based Measures}}} \\
\midrule
Simple Matching & Sokal-Michener & $\frac{a+d}{n}$ & Total agreement proportion; treats positive and negative matches equally & \\
\addlinespace[0.5ex]
Rogers-Tanimoto & & $\frac{a+d}{a+d+2(b+c)}$ & Penalized matching coefficient; reduces weight of disagreements & \\
\addlinespace[0.5ex]
Hamming Similarity & & $1-\frac{b+c}{n}$ & Complement of normalized edit distance; symmetric disagreement penalty & \\
\midrule
\multicolumn{5}{l}{\textit{\textbf{Correlation and Covariance Measures}}} \\
\midrule
Phi Coefficient & Matthews CC, Pearson & $\frac{ad-bc}{\sqrt{(a+b)(c+d)(a+c)(b+d)}}$ & Tetrachoric correlation for binary data; ranges [-1,1]; measures linear association & \\
\addlinespace[0.5ex]
Yule's Q & Goodman-Kruskal $\gamma$ & $\frac{ad-bc}{ad+bc}$ & Ordinal association measure; emphasizes concordant vs. discordant pairs & \\
\addlinespace[0.5ex]
Kendall's $\tau_b$ & & $\frac{ad-bc}{\sqrt{(n_c-T_x)(n_c-T_y)}}$ & Rank correlation with tie correction; robust to marginal distributions & \\
\addlinespace[0.5ex]
Cramér's V & & $\sqrt{\frac{\chi^2}{n}}$ & Chi-square based association; standardized effect size measure & \\
\midrule
\multicolumn{5}{l}{\textit{\textbf{Proportional Reduction in Error}}} \\
\midrule
Cohen's $\kappa$ & & $\frac{p_o - p_e}{1 - p_e}$ & Chance-corrected agreement; accounts for marginal probability agreement & \\
\addlinespace[0.5ex]
Adjusted Rand Index & & $\frac{\binom{n}{2}(P+Q) - [(P+R)(P+S)+(S+T)(R+T)]}{\binom{n}{2}^2 - [(P+R)(P+S)+(S+T)(R+T)]}$ & Clustering agreement corrected for chance; extended kappa for partitions & \\
\midrule
\multicolumn{5}{l}{\textit{\textbf{Information-Theoretic Measures}}} \\
\midrule
Mutual Information & & $I(X;Y) = H(X) + H(Y) - H(X,Y)$ & Shared information content; measures statistical dependence reduction in uncertainty & \\
\addlinespace[0.5ex]
Normalized MI & Symmetric Uncertainty & $\frac{2I(X;Y)}{H(X)+H(Y)}$ & Standardized mutual information; bounded [0,1]; symmetric measure & \\
\midrule
\multicolumn{5}{l}{\textit{\textbf{Psychometric Scaling Measures}}} \\
\midrule
Loevinger's $H_i$ & & $\frac{P_{ij} - P_i P_j}{\max(P_{ij}) - P_i P_j}$ & Scalability coefficient; measures departure from independence relative to maximum possible & \\
\addlinespace[0.5ex]
Standardized $Z_i$ & & $\frac{H_i}{\text{SE}(H_i)}$ & Standardized scalability; provides significance testing for monotone homogeneity & \\
\addlinespace[0.5ex]
2PL Discrimination & IRT $a$-parameter & $a_i = \frac{1.7\phi_{ij}}{\sqrt{1-\phi_{ij}^2}}$ & Item discrimination parameter; relates tetrachoric correlation to IRT slope & \\
\bottomrule
\end{tabular}%
}
\begin{tablenotes}
\small
\item Note: Contingency table notation: $a$ = both items positive, $b$ = item $i$ positive/$j$ negative, $c$ = item $i$ negative/$j$ positive, $d$ = both items negative, $n$ = total sample size. $P_i$, $P_j$ denote marginal probabilities; $P_{ij}$ denotes joint probability.
\end{tablenotes}
\end{table}

\begin{table}[htbp]
\centering
\caption{Specialized Binary Similarity Measures for Item Response Applications}
\label{tab:specialized_measures}
\resizebox{\textwidth}{!}{%
\begin{tabular}{>{\raggedright}p{2.8cm} >{\raggedright}p{2.2cm} >{\centering}p{4.5cm} >{\raggedright}p{5.8cm} p{1.2cm}}
\toprule
\textbf{Measure} & \textbf{Aliases} & \textbf{Formula} & \textbf{Description \& Key Features} & \textbf{Refs} \\
\midrule
\multicolumn{5}{l}{\textit{\textbf{Expectation-Adjusted Measures}}} \\
\midrule
Forbes Coefficient & & $\frac{na}{(a+b)(a+c)}$ & Ratio of observed to expected co-occurrence; sensitive to rare item pairs & \\
\addlinespace[0.5ex]
Tarwid Index & & $\frac{na - (a+b)(a+c)}{na + (a+b)(a+c)}$ & Standardized deviation from independence; symmetric about zero & \\
\addlinespace[0.5ex]
McConnaughey & & $\frac{a^2 - bc}{\sqrt{(a+b)(a+c)}}$ & Squared overlap relative to geometric mean; emphasizes strong associations & \\
\midrule
\multicolumn{5}{l}{\textit{\textbf{Monotonicity and Scalability}}} \\
\midrule
Symmetric $H$ & Mokken $H_{ij}$ & $1 - \frac{b+c}{2P_i(1-P_j) + 2P_j(1-P_i)}$ & Symmetric scalability index; measures violation of monotone homogeneity & \\
\addlinespace[0.5ex]
Isotonic $R^2$ & Unsigned Monotone & $\phi^2$ & Squared correlation; measures strength ignoring direction of association & \\
\addlinespace[0.5ex]
Signed Isotonic & Monotone $R^2$ & $\text{sign}(ad-bc) \cdot \phi^2$ & Directional monotone association; preserves sign of relationship & \\
\midrule
\multicolumn{5}{l}{\textit{\textbf{Robust and Weighted Measures}}} \\
\midrule
Baroni-Urbani-Buser & & $\frac{\sqrt{ad} + a}{\sqrt{ad} + a + b + c}$ & Includes negative matches in geometric weighting; robust to marginal asymmetry & \\
\addlinespace[0.5ex]
Yule's $W$ & & $\frac{\sqrt{ad} - \sqrt{bc}}{\sqrt{ad} + \sqrt{bc}}$ & Square-root stabilized odds ratio; reduces influence of extreme cell counts & \\
\addlinespace[0.5ex]
Stiles Index & & $\log_{10}\left(\frac{n(|ad-bc|-n/2)^2}{(a+b)(c+d)(a+c)(b+d)}\right)$ & Yates-corrected chi-square; provides continuity correction for small samples & \\
\midrule
\multicolumn{5}{l}{\textit{\textbf{Distance-Based Transformations}}} \\
\midrule
Hellinger Similarity & & $1 - \sqrt{\frac{1}{2}\left(\frac{b}{a+b} + \frac{c}{a+c}\right)}$ & Probabilistic distance measure; derived from Hellinger distance between distributions & \\
\addlinespace[0.5ex]
Austin-Colwell & & $\frac{2}{\pi}\arcsin\left(\sqrt{\frac{a+d}{n}}\right)$ & Arcsine-transformed agreement; variance-stabilizing transformation & \\
\addlinespace[0.5ex]
Pattern Difference & & $1 - \frac{4bc}{n^2}$ & Shape-based similarity; measures configurational rather than marginal differences & \\
\bottomrule
\end{tabular}%
}
\begin{tablenotes}
\small
\item Note: These measures address specific psychometric challenges including marginal heterogeneity, sample size sensitivity, and monotonicity assumptions in item response modeling. Selection depends on theoretical assumptions about item relationships and desired mathematical properties.
\end{tablenotes}
\end{table}

\begin{table}[ht!]
\resizebox{\textwidth}{!}{%
\begin{tabular}{@{} lllp{6.8cm} @{}} 
\toprule
\textbf{Name}     & \textbf{Aliases} & \textbf{Formula} & \textbf{Salient Features \& Use Case} \\
\midrule

\multicolumn{4}{l}{\textbf{I. Intersection-Based Measures (Joint Positive Occurrence)}} \\
\addlinespace[0.2em]
Dice Precision & $S_P$ (Precision) & $\frac{a}{a+b}$ & Proportion of $X=1$ also $Y=1$; insensitive to co-absences, \textit{directional}. \\
Dice Recall    & $S_R$ (Recall)    & $\frac{a}{a+c}$ & Proportion of $Y=1$ also $X=1$; equally directional. \\
Jaccard        & (Tanimoto)        & $\frac{a}{a+b+c}$ & Fraction of total positives shared; ignores double-zeros; popular in ecology/text. \\
Weighted Jaccard &                & $\frac{3a}{3a + b + c}$ & Increased weight on matches; strengthens intersection focus. \\
Russell-Rao    &                  & $\frac{a}{n}$   & Absolute fraction of joint positives; marginal coverage emphasized, sparse for rare items. \\
Sokal-Sneath 1 &                  & $\frac{a}{a+2(b+c)}$ & Penalizes discord more than Jaccard; stricter similarity. \\
Braun-Blanquet &                  & $\frac{a}{\max(a+b,\,a+c)}$ & Ratio of joint positives to maximum positive marginal; overlap normalization. \\
Mountford      &                  & $\frac{2a}{a(b+c)+2bc}$ & Sensitive to shared presence among rare items. \\
Sorgenfrei     &                  & $\frac{a^2}{(a+b)(a+c)}$ & Squared joint positive, normalized for expected chance; rare-item association. \\
Ochiai/ Cosine &                  & $\frac{a}{\sqrt{(a+b)(a+c)}}$ & Geometric mean scaling; adjusts for marginal frequencies; popular in text mining. \\
\addlinespace[0.4em]

\multicolumn{4}{l}{\textbf{II. Agreement (Matching) Measures}} \\
\addlinespace[0.2em]
Simple Matching (SMC) &        & $\frac{a+d}{n}$ & Overall agreement; considers both co-presence and co-absence. \\
Sokal-Sneath 3        &        & $\frac{a+d}{b+c}$ & Ratio of total agreement to disagreement; unbounded above. \\
Gower                &         & $\frac{a+d}{\sqrt{(a+b)(a+c)(b+d)(c+d)}}$ & Adjusts for marginals; robust for unbalanced data. \\
Rogot-Goldberg        &        & $\frac{a}{2a+b+c} + \frac{d}{2d+b+c}$ & Symmetrized precision for $a$ and $d$; accentuates strong matches. \\
\addlinespace[0.4em]

\multicolumn{4}{l}{\textbf{III. Distance and Dissimilarity-Derived Measures}} \\
\addlinespace[0.2em]
Normalized Euclidean  &        & $1 - \frac{\sqrt{b+c}}{\sqrt{n}}$ & Transforms Hamming/\(L_2\) distance to similarity; interpretable as spatial proximity. \\
Canberra              &        & $1-\frac{1}{2}\left( \frac{b}{a+b}+\frac{c}{a+c} \right) $ & Emphasizes mismatches when marginals are small. \\
Size Difference       &        & $1 - \frac{(b+c)^2}{n^2}$ & Quadratic disagreement penalty; accentuates outlier differences. \\
\addlinespace[0.4em]

\multicolumn{4}{l}{\textbf{IV. Covariance, Correlation, and Dependency Measures}} \\
\addlinespace[0.2em]
\textit{Determinant}  &        & $ad - bc$ & Raw cross-patterning between pairs; basis for several associations below. \\
Yule's Q (Goodman-Kruskal $\gamma$) & & $\frac{ad-bc}{ad + bc}$ & Scaling from perfect agreement ($1$) to inversion ($-1$); extreme-focus. \\
Phi / Matthews Correlation  &   & $\frac{ad-bc}{\sqrt{(a+b)(a+c)(b+d)(c+d)}}$ & Pearson for binaries; $[-1,1]$ range; symmetric; interpretable as "latent" correlation. \\
Cohen's Kappa         &        & $\frac{P_o - P_e}{1 - P_e}$ & Chance-corrected agreement; robust to marginal imbalance. \\
Kendall's Tau-a       &        & $\frac{ad-bc}{n(n-1)/2}$ & Concordance-discordance scaled for pairs; rank-based. \\
Tetrachoric Approx.   &        & $\cos\left(\frac{\pi}{1 + \sqrt{Odds\,Ratio}}\right)$ & Approximates correlation between dichotomized normals. \\
\addlinespace[0.4em]

\multicolumn{4}{l}{\textbf{V. Information-Theoretic Measures}} \\
\addlinespace[0.2em]
Mutual Information    &        & $MI(X;Y) = H(X)+H(Y)-H(X,Y)$ & Nonlinear association; accounts for total and joint uncertainty. \\
Theil's $U$ (Uncertainty coefficient) & & $\frac{H(Y) - H(Y|X)}{H(Y)}$ & Proportion of $Y$'s entropy explained by $X$; highly directional, bounded $[0,1]$. \\
\addlinespace[0.4em]

\multicolumn{4}{l}{\textbf{VI. Cluster and Proportional Reduction in Error Indices}} \\
\addlinespace[0.2em]
Rand Index           &         & $\frac{A+ad}{\tbinom{n}{2}}$ & Agreement over all item pairs; adjusts for cluster structure and chance. \\
Goodman-Kruskal Lambda &        & See note$^a$ & Proportion by which knowing one variable reduces error in predicting the other; direction matters. \\
\bottomrule
\end{tabular}
}
\medskip
\raggedright
\footnotesize
\textbf{Notes.} Notation for marginal and joint counts: $a$, $b$, $c$, $d$ as in the text; $n = a+b+c+d$.
Where other symbols appear: $P_o$ is observed agreement, $P_e$ is expected agreement by chance, $Odds\,Ratio = \frac{ad}{bc}$, $H(\cdot)$ is (Shannon) entropy, and $A$ is the number of pairs clustered together in both variables. $^a$ See Supplement for full Goodman-Kruskal formulae and context.

\end{table}

%% file: apx/measure_theory.tex
\section{Formal Treatment of Isotonic Regression}
This appendix builds the mathematical relationships of isotonic regression in the context of scalability and bad item detection.

\subsection{The Probability Space}

We begin by defining the measure-theoretic foundation. Let $(\Omega, \mathcal{F}, P)$ be a probability space, where:
\begin{itemize}
    \item $\Omega$ is the sample space (a set of outcomes).
    \item $\mathcal{F}$ is a $\sigma$-algebra of events (a set of subsets of $\Omega$).
    \item $P$ is a probability measure on $\mathcal{F}$.
\end{itemize}

Let $X$ and $Y$ be two real-valued random variables, which are measurable functions from our probability space to the real numbers:
\begin{itemize}
    \item $X: \Omega \to \mathbb{R}$
    \item $Y: \Omega \to \mathbb{R}$
\end{itemize}

The expectation of a random variable $Z$ (or any integrable function $g(X,Y)$) is its Lebesgue integral with respect to the measure $P$:
$E[Z] := \int_{\Omega} Z(\omega) \,dP(\omega)$

We assume that $X$ and $Y$ have finite second moments, i.e., $E[X^2] < \infty$ and $E[Y^2] < \infty$. This ensures that their variances are well-defined and finite.
\begin{itemize}
    \item $\text{Var}(X) = E[(X - E[X])^2] = \int_{\Omega} (X(\omega) - E[X])^2 \,dP(\omega)$
    \item $\text{Var}(Y) = E[(Y - E[Y])^2] = \int_{\Omega} (Y(\omega) - E[Y])^2 \,dP(\omega)$
\end{itemize}

\subsection{Signed Isotonic (Monotonic) Regression}

Isotonic regression finds the best-fitting monotonic function. Unlike linear regression, we search over the entire space of monotonic functions to minimize the mean squared error (MSE).

We define the two sets of monotonic, measurable functions:
\begin{itemize}
    \item The set of non-decreasing (isotonic) functions: $\mathcal{M}^{\uparrow} = \{f: \mathbb{R} \to \mathbb{R} \mid f \text{ is measurable and } x_1 \le x_2 \implies f(x_1) \le f(x_2)\}$
    \item The set of non-increasing (antitonic) functions: $\mathcal{M}^{\downarrow} = \{f: \mathbb{R} \to \mathbb{R} \mid f \text{ is measurable and } x_1 \le x_2 \implies f(x_1) \ge f(x_2)\}$
\end{itemize}

\subsection{Signed $R^2$ for $Y$ regressed on $X$}

First, we consider the regression of $Y$ on $X$. We must find the best monotonic fit, which could be either non-decreasing or non-increasing.

1.  Find the best non-decreasing fit ($f^{\uparrow}_{Y|X}$): This function minimizes the MSE over all functions in $\mathcal{M}^{\uparrow}$.
    $f^{\uparrow}_{Y|X} := \arg\min_{f \in \mathcal{M}^{\uparrow}} E[(Y - f(X))^2] = \arg\min_{f \in \mathcal{M}^{\uparrow}} \int_{\Omega} (Y(\omega) - f(X(\omega)))^2 \,dP(\omega)$
    Let the resulting minimum MSE be $\text{MSE}^{\uparrow}_{Y|X} = E[(Y - f^{\uparrow}_{Y|X}(X))^2]$.

2.  Find the best non-increasing fit ($f^{\downarrow}_{Y|X}$): This function minimizes the MSE over all functions in $\mathcal{M}^{\downarrow}$.
    $f^{\downarrow}_{Y|X} := \arg\min_{f \in \mathcal{M}^{\downarrow}} E[(Y - f(X))^2] = \arg\min_{f \in \mathcal{M}^{\downarrow}} \int_{\Omega} (Y(\omega) - f(X(\omega)))^2 \,dP(\omega)$
    Let the resulting minimum MSE be $\text{MSE}^{\downarrow}_{Y|X} = E[(Y - f^{\downarrow}_{Y|X}(X))^2]$.

3.  Determine the Direction and the Signed R²: The direction of the monotonicity is determined by which fit is better (i.e., has a lower MSE). We define a sign, $\sigma_{Y|X}$, based on this comparison.
    $\sigma_{Y|X} := \begin{cases} +1 & \text{if } \text{MSE}^{\uparrow}_{Y|X} \le \text{MSE}^{\downarrow}_{Y|X} \\ -1 & \text{if } \text{MSE}^{\uparrow}_{Y|X} > \text{MSE}^{\downarrow}_{Y|X} \end{cases}$

    The overall best monotonic MSE is $\min(\text{MSE}^{\uparrow}_{Y|X}, \text{MSE}^{\downarrow}_{Y|X})$. The corresponding unsigned $R^2$ is:
    $R^2_{Y|X} = 1 - \frac{\min(\text{MSE}^{\uparrow}_{Y|X}, \text{MSE}^{\downarrow}_{Y|X})}{\text{Var}(Y)}$

    The signed R² for Y on X, which we denote $S^2_{Y|X}$, is the product of the sign and the unsigned R²:
    $S^2_{Y|X} := \sigma_{Y|X} \cdot \left(1 - \frac{\min(E[(Y - f^{\uparrow}_{Y|X}(X))^2], E[(Y - f^{\downarrow}_{Y|X}(X))^2])}{\text{Var}(Y)}\right)$

\subsection{Signed R² for X regressed on Y}

The process is perfectly symmetric. We now regress $X$ on $Y$. We need a new set of functions, which we'll call $g$, to avoid confusion.

1.  Find the best non-decreasing fit ($g^{\uparrow}_{X|Y}$):
    $g^{\uparrow}_{X|Y} := \arg\min_{g \in \mathcal{M}^{\uparrow}} E[(X - g(Y))^2]$
    $\text{MSE}^{\uparrow}_{X|Y} = E[(X - g^{\uparrow}_{X|Y}(Y))^2]$

2.  Find the best non-increasing fit ($g^{\downarrow}_{X|Y}$):
    $g^{\downarrow}_{X|Y} := \arg\min_{g \in \mathcal{M}^{\downarrow}} E[(X - g(Y))^2]$
    $\text{MSE}^{\downarrow}_{X|Y} = E[(X - g^{\downarrow}_{X|Y}(Y))^2]$

3.  Determine the Direction and the Signed R²:
    $\sigma_{X|Y} := \begin{cases} +1 & \text{if } \text{MSE}^{\uparrow}_{X|Y} \le \text{MSE}^{\downarrow}_{X|Y} \\ -1 & \text{if } \text{MSE}^{\uparrow}_{X|Y} > \text{MSE}^{\downarrow}_{X|Y} \end{cases}$

    The signed R² for X on Y, denoted $S^2_{X|Y}$, is:
    $S^2_{X|Y} := \sigma_{X|Y} \cdot \left(1 - \frac{\min(E[(X - g^{\uparrow}_{X|Y}(Y))^2], E[(X - g^{\downarrow}_{X|Y}(Y))^2])}{\text{Var}(X)}\right)$

\subsection{Final Notation for the Association Measure}

The final association measure is the mean of the two signed R² values. Let's call this measure $\rho_m^2(X, Y)$, where the subscript 'm' stands for monotonic.

The final mathematical notation is:

$\rho_m^2(X, Y) = \frac{1}{2} \left( S^2_{Y|X} + S^2_{X|Y} \right)$

where:

$S^2_{Y|X} = \text{sgn}(\text{MSE}^{\downarrow}_{Y|X} - \text{MSE}^{\uparrow}_{Y|X}) \cdot \left(1 - \frac{\min(\text{MSE}^{\uparrow}_{Y|X}, \text{MSE}^{\downarrow}_{Y|X})}{\text{Var}(Y)}\right)$

$S^2_{X|Y} = \text{sgn}(\text{MSE}^{\downarrow}_{X|Y} - \text{MSE}^{\uparrow}_{X|Y}) \cdot \left(1 - \frac{\min(\text{MSE}^{\uparrow}_{X|Y}, \text{MSE}^{\downarrow}_{X|Y})}{\text{Var}(X)}\right)$

and the components are defined using Lebesgue integrals over the probability space $(\Omega, \mathcal{F}, P)$ as follows:

   \begin{itemize}
       \item $\text{MSE}^{\uparrow}_{Y|X} = \inf_{f \in \mathcal{M}^{\uparrow}} \int_{\Omega} (Y(\omega) - f(X(\omega)))^2 \,dP(\omega)$
       \item $\text{MSE}^{\downarrow}_{Y|X} = \inf_{f \in \mathcal{M}^{\downarrow}} \int_{\Omega} (Y(\omega) - f(X(\omega)))^2 \,dP(\omega)$
       \item $\text{MSE}^{\uparrow}_{X|Y} = \inf_{g \in \mathcal{M}^{\uparrow}} \int_{\Omega} (X(\omega) - g(Y(\omega)))^2 \,dP(\omega)$
       \item $\text{MSE}^{\downarrow}_{X|Y} = \inf_{g \in \mathcal{M}^{\downarrow}} \int_{\Omega} (X(\omega) - g(Y(\omega)))^2 \,dP(\omega)$
       \item $\text{Var}(Y) = \int_{\Omega} \left(Y(\omega) - \int_{\Omega} Y(\omega') dP(\omega')\right)^2 dP(\omega)$
       \item $\text{Var}(X) = \int_{\Omega} \left(X(\omega) - \int_{\Omega} X(\omega') dP(\omega')\right)^2 dP(\omega)$
   \end{itemize}


$\rho_{miso}^2(X, Y) =  \text{sgn}_{mono} \cdot \left(2 - \frac{\underset{g \in \mathcal{M}}{\operatorname{argmin}}(\int_{\Omega} (X(\omega) - g(Y(\omega)))^2 \,dP(\omega))}{\int_{\Omega} \left(Y(\omega) - \int_{\Omega} Y(\omega') dP(\omega')\right)^2 dP(\omega)} - \frac{\underset{f \in \mathcal{M}}{\operatorname{argmin}} (\int_{\Omega} (Y(\omega) - f(X(\omega)))^2 \,dP(\omega))}{\int_{\Omega} \left(X(\omega) - \int_{\Omega} X(\omega') dP(\omega')\right)^2 dP(\omega)}\right)$

   $\mathcal{M}^{\uparrow}$ and $\mathcal{M}^{\downarrow}$ are the sets of non-decreasing and non-increasing measurable functions, respectively.
   The use of `inf` (infimum) is slightly more rigorous than `min` (minimum) as the minimum might not be achieved by a specific function within the set, though in the context of these $L^2$ projections, it is. The `argmin` notation used earlier implies this existence.
   The sign function `sgn` is defined as $\text{sgn}(z)=+1$ for $z \ge 0$ and $\text{sgn}(z)=-1$ for $z < 0$, which matches our definition of $\sigma$.